\documentclass[11pt, oneside]{article} 
\usepackage{amsmath, amsthm, amssymb, calrsfs, wasysym, verbatim, bbm, color, graphics, geometry}

\usepackage{graphicx}
\usepackage{url}
\usepackage{enumitem}
\usepackage{setspace}
\usepackage{hyperref}

\title{New Horizons for Psi \\ Studying fundamental fields with numerical relativity}
\author{Katy Clough}
\date{July 2024}

\begin{document}

\maketitle
\tableofcontents

\newpage

\section{Overview of the course}

This set of notes was designed to accompany two hours of lectures and practical exercises at the New Horizons for Psi workshop \url{https://strong-gr.com/new-horizons-for-psi}, entitled ``Studying fundamental fields with numerical relativity''. 

Numerical relativity (NR) is a tool used to help understand the behaviours of metric and matter fields in dynamical, strong gravity situations. 
It has been used to study a range of situations involving fundamental fields, including superradiance, modified gravity, dynamical friction, dark matter accretion and early universe cosmology. We will hear more about these in the workshop. The purpose of this course is to provide you with some background and hands on experience in numerical relativity that will help you to better understand the possibilities and limitations provided by this tool.

We will use the code \texttt{engrenage}. 
This is a spherically symmetric code designed for teaching NR that uses the reference metric framework and a dynamical gauge. The code includes a scalar field obeying the Klein Gordon (KG) equation for a minimally coupled spin 0 field as the matter source of the metric curvature. The version of the code we will use in the course can be found here: \\
\url{https://github.com/GRTLCollaboration/engrenage/tree/NewHorizonsForPsi}.\\
The code is based on the formalism in the papers \cite{Ruchlin:2017com,Baumgarte:2012xy,Brown:2009dd} and many people have contributed to its development - please see the acknowledgements for more details.

My assumption for these notes is that people know GR and may have come across the ADM decomposition, but don't know numerical methods or numerical relativity itself.
For those who are already familiar with numerical methods and the ADM decomposition you can skip directly to the exercises and try to go as far as you can with them. There is more there than you can do in 2 hours, so feel free to pick and choose exercises you feel most motivated to try, they can all be completed independently of each other. 
For more details see the papers above or the key NR texts \cite{Alcubierre:2008co, Baumgarte:2021skc,Baumgarte:2010ndz,Shibata_book,Gourgoulhon:2007ue} (\cite{Baumgarte:2021skc} is particularly good for beginners). The wiki for the course also provides relevant background material: \\
\url{https://github.com/GRTLCollaboration/engrenage/wiki/New-Horizons-for-Psi-workshop}. 

\textit{Disclaimer:} The \texttt{engrenage} code was not designed to be a good example of optimised python usage. The goal was to write a code where some non trivial physical examples could be studied and users could get an overview of the different parts of a numerical relativity code with a dynamical gauge, without the obfuscation that arises due to the optimisation and HPC requirements of a typical research code like its parent GRChombo \cite{Andrade:2021rbd}. However, thanks to input from others it now has quite good performance and I have used it successfully for several testing and research projects. Masters students in particular benefit from the simple python format, which means they can get started quickly with implementing actual physics.\footnote{The original working name of the code was \textit{BabyGRChombo}, \texttt{engrenage} (a system of gears) was chosen because it captures the idea of being able to see the connections between different parts of the code. As we are in Portugal we should perhaps say \texttt{engrenagem}.}.

\newpage
\section{Review of numerical methods}

The knowledge in this section is essential for understanding how the code works. It its not a rigorous introduction to numerical methods, and just aims to communicate the key concepts sufficiently to be able to get started. Those already familiar with numerical methods can skip it.

\subsection{Basic principles of numerical methods}

If you are an analytic person, you probably consider ``solving'' an equation to be a global process. Given an equation like
\begin{equation}
    \frac{du}{dt} = -3u ~,
\end{equation}
You immediately see that the solution for \textit{all time} is 
\begin{equation}
    u = Ce^{-3t}
\end{equation}
and then as a final step you may use a single initial condition relating $t$ and $u$ at some time to fix the constant. 

This is conceptually very different to the numerical method of solution. Here we start with the initial condition, e.g. $u(t=0) = 1$ and then integrate the equation in a discrete way to get the next value, that is we do something like
\begin{verbatim}
    u_old = 1.0
    t_old = 0.0
    dudt = -3.0 * u_old
    dt = 0.1
    u_new = u_old + dudt * dt
    t_new = t_old + dt
\end{verbatim}
Now we can replace the old values with the new values and repeat to populate the values for larger $t$ until we get bored or run out of memory. If we plot the result, this will show us the shape of the solution, but we still won't have the analytic form for it (although we might be able to fit it with some function if it is simple).

Whilst this may seem like the ``level-zero'' way to solve an equation of motion, it is extremely powerful as a method and moreover, something that can give you better physical intuition for how systems evolve. GR is (ignoring some extensions) a \textit{local} theory, and this method forces you to think about the evolution at each point in space and time and its dependence on information from other parts of space and time. ``Nature does not care about our mathematical difficulties; it integrates numerically'' (to paraphrase Einstein!).

This trivial example also nicely illustrates most of the key advantages and disadvantages of numerical relativity:
\begin{itemize}
    \item \textbf{Accuracy and error:} Unless we take a very small step size, we are not going to get a very accurate result. The algorithm described above is Euler's method, and the global error scales as $dt$. Fortunately there are other methods that make the integration more accurate and efficient, as we will see below, but this problem always persists at some level. An NR solution is not exact, and always includes some error, but that error should be quantifiable, so we know at what level to trust the result.
    \item \textbf{No general solution:} Because I am numerically solving from a single initial condition, I get one solution for that initial condition and nothing more. If my equation also includes unknown parameters (e.g. the mass of a scalar field, or a coupling constant), I also need to assign them a numerical value, so that I can evaluate the results at each step. So again my result is specific to the chosen parameters. This is a major limitation for NR, where each individual simulation can be very costly. The trick of using NR to understand some physics is to be very clear about the question you want to ask, and how you are going to test it. Just blindly running a huge parameter scan usually won't help you to learn anything more broadly applicable, and is a waste of resources.
    \item \textbf{Analytic intractability} Because I evaluate the expressions numerically, it (almost) doesn't matter how complicated they are. I can do something like 
    \begin{equation}
        \frac{du}{dt} = 3 t^2 e^{\sin(u)} ~,
    \end{equation}
    that would give an undergraduate student in an exam a panic attack, and a minor amendment to the algorithm above will give me the solution. This is the power of NR, because the GR equations are in most general cases analytically untractable. In the code you will see that the time derivatives for the metric fields look pretty ugly when written out in full, and they did take a while to debug, but conceptually it wasn't more work that adding the simpler Klein-Gordon equation.
    \item \textbf{Timescales:} A bit more subtle. But imagine that the equation contains a term like
    \begin{equation}
        \frac{du}{dt} = \sin(\omega_1 t) \cos(\omega_2 t) ~,
    \end{equation}
    where $\omega_1=1000$, and $\omega_2=0.001$. I am going to have to take very small steps $dt$ to make sure I resolve all the finer oscillations from the first high frequency scale, but I won't see the longer period ones unless I get to large $t$. This is going to mean a lot of steps. The conclusion is that problems that include two or more very different time or length scales are not usually well suited to numerical relativity\footnote{On the bright side, those are often the ones that are amenable to perturbative methods like Post Newtonian (PN) calculations, which we would much rather use if we can because of their generality.}.
\end{itemize}

\noindent \textit{Main message:} For most modern research problems, numerical methods are not a magic black box that can solve all problems. They need to be applied with judgement, ideally in parallel to analytic methods with a view to verifying their approximations and getting some semi-analytic handle on the problem. It is common to need to compromise on the exact physical setup, and reduce it to a representative toy model in order to make it more tractable. This creative aspect of numerical relativity is what I enjoy most - how can I distill my complex problem into the simplest case that can be studied numerically, but is still meaningful?

\subsection{How do I represent data on a computer?}

The solution data in the \texttt{engrenage} code represents a set of 2D functions $f(r,t)$ for each of the metric and matter variables. Computers only store discrete 1s and 0s, so how can I represent these functions, which should be continuous? 

\begin{figure}
\centering
\includegraphics[width=0.8\textwidth]{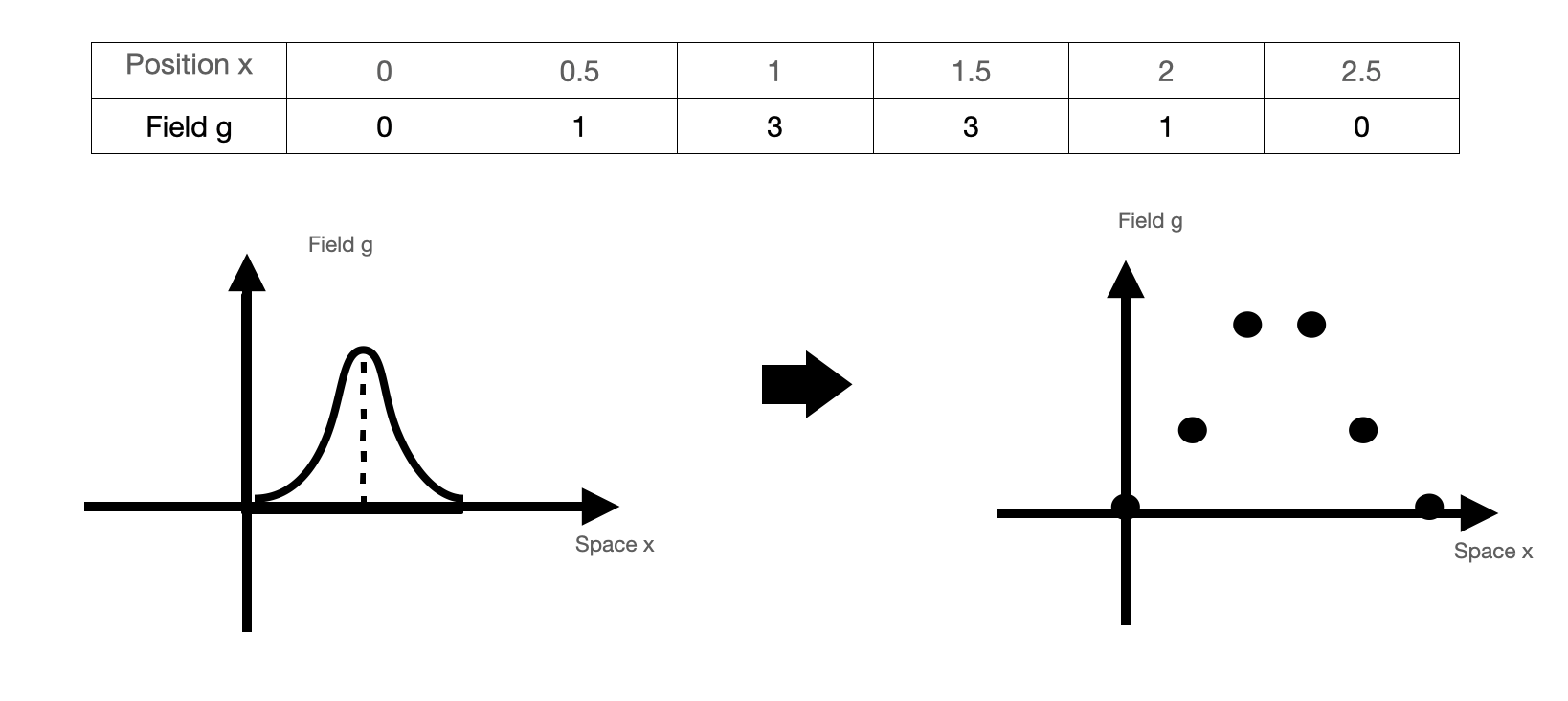}
\caption{A continuous function is represented as a series of values at discrete points. For multiple variables in 1+1D spaces (i.e., dependence on $r$ and $t$) we have a vector of spatial values like this for each variable, and this set of vectors (which I will refer to as the ``state vector'') exists for every time point.}
\label{fig-DiscreteData}
\end{figure}

If I knew more about the exact form of the solution, I could very efficiently store it with an appropriate set of basis functions, e.g. as a sum of sines and cosines with different frequencies. Unfortunately, we don't usually know the form of the solution ahead of time, and it won't usually be periodic, or even smooth (around a BH singularity, for example). For this reason, we often simply store the values at a set of discrete points defined on a uniform coordinate grid
\footnote{Some NR codes employ ``pseudospectral methods'' which use a basis of Chebyshev polynomials. These methods can be highly accurate and efficient, at the cost of somewhat greater complexity in the implementation.}. 
This is illustrated in Fig. \ref{fig-DiscreteData}.

\subsection{How do I take (spatial) derivatives of discrete data?}

\begin{figure}
\centering
\includegraphics[width=0.8\textwidth]{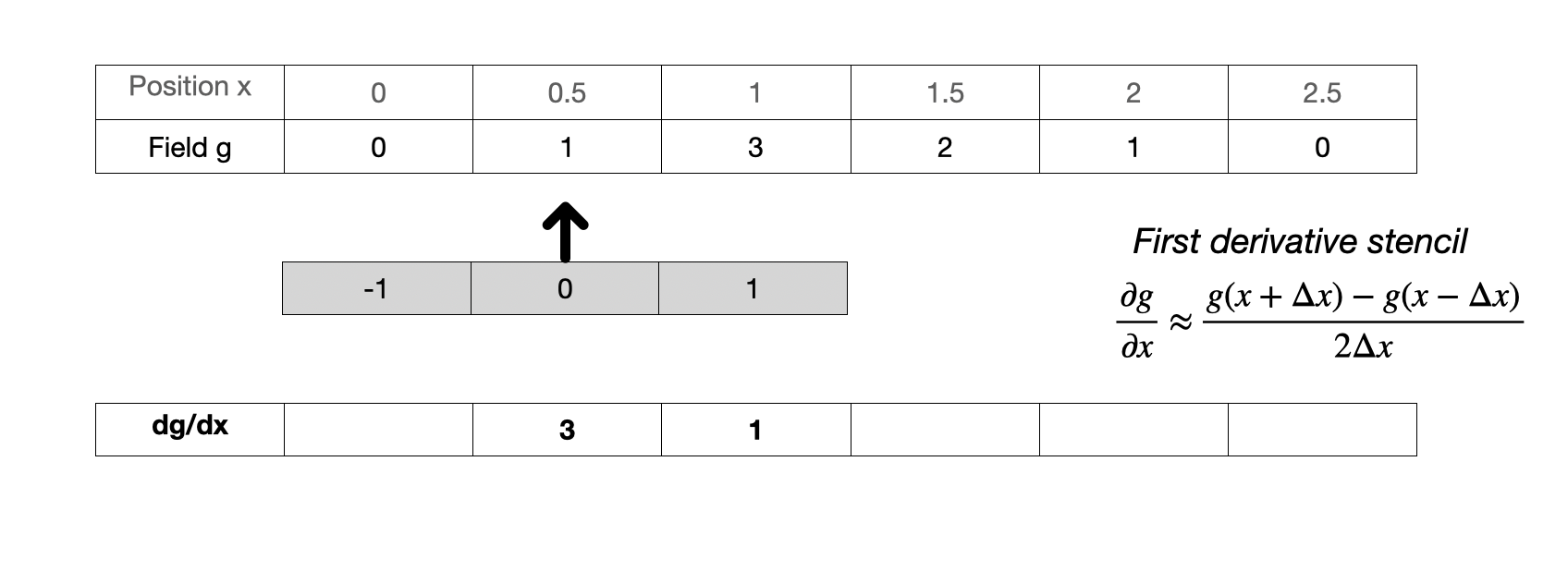}
\caption{The stencil is convolved with each point and its surrounding values on the grid. Note that the grid spacing here is $\Delta x = 0.5$. At the end cells (referred to in the code as ``ghost cells'') we need to impose boundary conditions or use one sided stencils.}
\label{fig-FDStencil}
\end{figure}

Now I have my data values at discrete points, but I will need to be able to take derivatives of it, since spatial derivatives appear in the equations of motion. 
A common way to do this (and that used by \texttt{engrenage}) is to use finite difference methods. 
This simply means that for the first derivative with respect to r, we make an approximation like
\begin{equation}
    \frac{\partial g}{\partial r} \approx \frac{(g(r+\Delta r) - g(r- \Delta r))}{2 \Delta r} ~.
\end{equation}
This gives us a ``finite difference stencil'' $[-1/(2\Delta r),~0,~ 1/(2\Delta r)]$, where the central value is the weighting of the current grid point, and the other two are those applied to the values of the variable either side. This stencil is then convolved with the function values across the grid to obtain a vector of values for the derivatives. This is illustrated in Fig. \ref{fig-FDStencil}.

By including more surrounding points (by increasing the complexity of the ``stencil'') we can get higher accuracy and the cost of somewhat greater complexity. The coefficients of the stencil can be obtained from fitting a Lagrange polynomial to the points and evaluating its derivatives in terms of the function values and their spacing
\footnote{Note that a very useful reference is the online MIT finite difference calculator 
\url{https://web.media.mit.edu/~crtaylor/calculator.html} but this only works for uniformly spaced grids.}.

\begin{figure}
\centering
\includegraphics[width=0.8\textwidth]{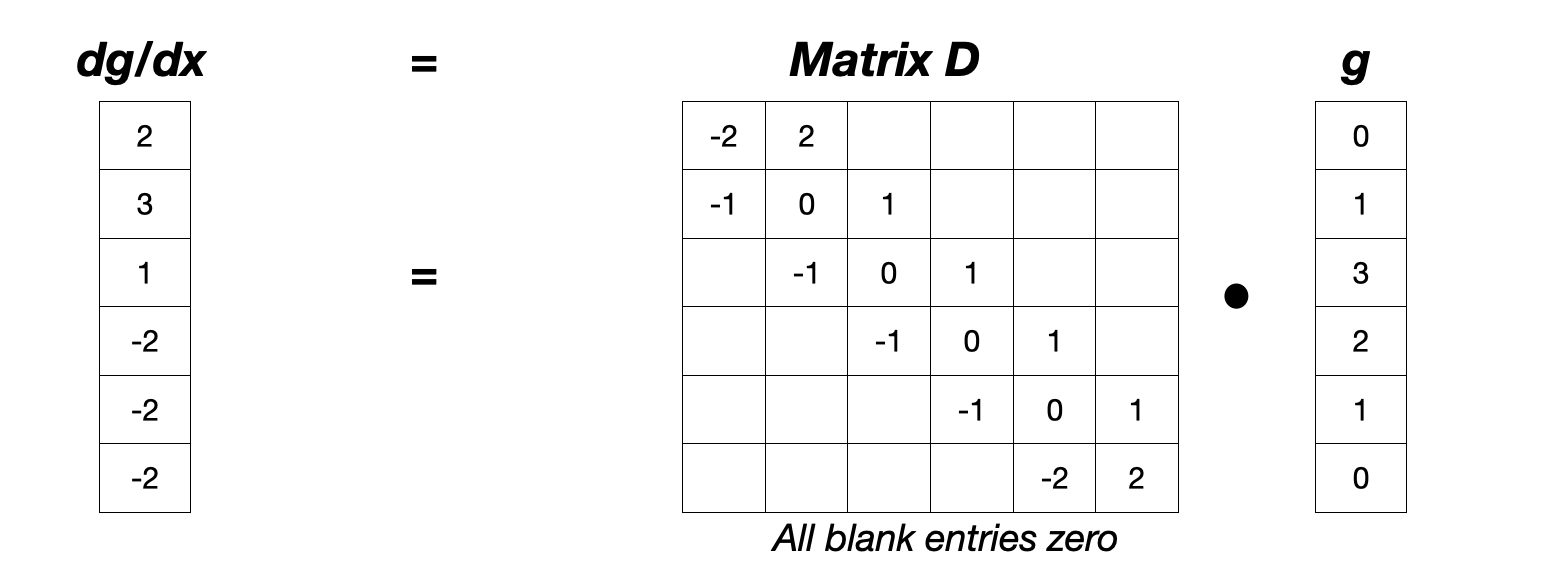}
\caption{The finite difference stencils can be packaged into a matrix that is applied to the whole vector of values for a given variable at a certain time, to return a vector of the values of the derivatives at each point.}
\label{fig-FDMatrix}
\end{figure}

These stencils can be packaged into a matrix that is applied to the whole vector of values for a given variable at a certain time, to return the values of the derivatives at each point. You will see this in the code in the file \verb|rhsevolution.py|, and it is illustrated schematically in Fig. \ref{fig-FDMatrix}.

\subsection{How do I integrate in time?}

We make use of the highly optimised python integrator \verb|solve_ivp| to perform the time integration of our solution. This essentially just takes the initial state, and given a function which computes the ``right hand side'' (often denoted ``rhs'', and referring to the time derivative $\partial_t g$) does something like:
\begin{equation}
    g(t + \Delta t) = g(t) + \partial_t g ~ \Delta t ~,
\end{equation}
using an algorithm similar to the one above. Again in practise it uses a higher order method (RK4, but checking and adjusting the timestep using RK5 to control the error), but you don't need to worry too much about the detail of that. 

Note that \verb|solve_ivp| works on a single vector of values, so we need to package up all the vectors for each variable (each containing the values at all the $r_i$ points) into one long state vector that contains an ordered set of all the variables at all the points $r_i$ on the spatial grid at a fixed time $t$. The solution data is a set of these state vector at a series of user-requested times (note that these times do not have any influence on the time stepping of the integrator).

Another technicality is that we actually have a second time derivative for the metric, since the Einstein equation is a non linear version of the wave equation
\begin{equation}
    \frac{\partial^2 g}{\partial t^2} - \frac{\partial^2 g}{\partial x^2} = ~ {\rm Source ~and~ non~ linear~ terms}~.
\end{equation}
The standard numerical trick is to define the first derivative as an independent variable $K$ and then split the second order equation into two coupled first order ones, that is, 
\begin{equation}
    \frac{\partial K}{\partial t} = \frac{\partial^2 g}{\partial x^2} + ~ {\rm Source}
\end{equation}
and
\begin{equation}
    \frac{\partial g}{\partial t} = K
\end{equation}
This splitting the second order (in time) Einstein equation into two first order equations of motion is naturally achieved in the ADM decomposition, which is discussed in the following section. It is also obvious from this decomposition that we not only need the initial condition for $g$, but also for its time derivative $K$, in order to generate a solution.

\newpage
\section{Review of numerical relativity} 

The standard approach of NR is to formulate and solve for solutions of GR as a $3+1$D Cauchy problem, using the numerical approach described in the previous section. This is a very physical approach - when I want to solve the Einstein equation numerically, the usual scenario is that I know or guess some initial condition on a spatial hypersurface (some black holes are orbiting each other), and want to find out ``what happens next'' (do they merge? What are the gravitational waves emitted?). That is, I want to evolve the spatial slice forward in time. Given the global hyperbolicity of the Einstein equations, this is in principle a tractable problem. If one knows the metric on a spatial hyperslice and its derivatives as one moves ``off'' the slice, that should be enough to populate the solution in the rest of the spacetime. What follows is a brief overview of the key points, and I refer the reader to the standard NR texts \cite{Alcubierre:2008co, Baumgarte:2021skc,Baumgarte:2010ndz,Shibata_book,Gourgoulhon:2007ue} for more details.

\subsection{Projections to 3+1D and the ADM formalism}

There exists a ``natural'' decomposition of the Einstein equations that is well motivated from both the Lagrangian and geometric approaches - \textit{the ADM (Arnowitt Deser Misner) decomposition}. This is the starting point for NR but sadly not the end point, since it is numerically unstable. We will not have time to discuss stability issues in much detail, but it is the main reason that, despite the ADM decomposition being published in 1959, NR has only become widely used in the last decade or two.

Nevertheless, the ADM decomposition contains most of the really key NR concepts, and it is common for people to talk in terms of the ADM decomposition variables, even though they may use slightly different ones in their chosen stable numerical formulation.

One of the major challenges of the ADM formalism is to decide how to achieve the foliation of the 4D spacetime into a series of spatial hypersurface. In a general GR problem, there is no preferred time-like direction and, crucially, no global concept of time. This makes the problem of solving the Einstein equation numerically substantially different from normal Cauchy problems like solving for a fluid flow. The data on the initial 3 dimensional spatial hyperslice is evolved forward along a local time coordinate, with each point corresponding to an \textit{observer} who moves through the spacetime, rather than any fixed spatial point. The freedom to choose the path of these observers, the so-called \textit{gauge choice}, is one of the most difficult issues in NR, although clearly any physical quantity that we extract from the simulation should be independent of that choice.

\begin{figure}
\centering
\includegraphics[width=0.5\textwidth]{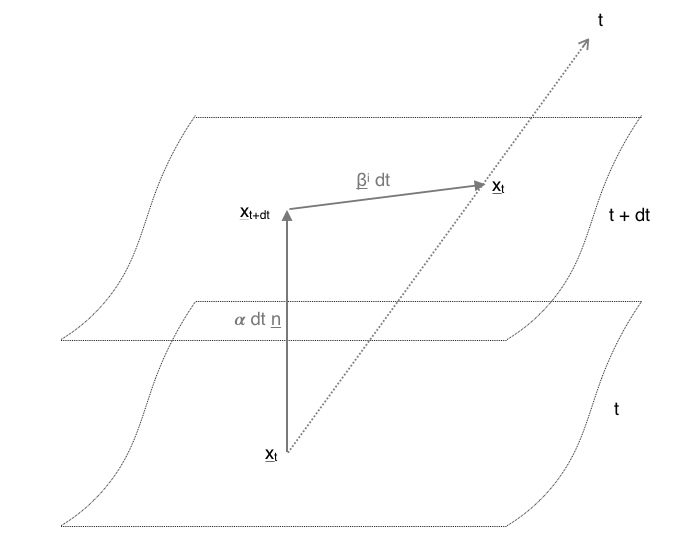}
\caption[$3+1$D Foliation of spacetime]{Foliation of a 4 dimensional spacetime into a 3 dimensional ``spatial'' hyperslice, and a ``time-like'' normal to that slice. The gauge variables - the lapse $\alpha$ and shift $\beta^i$ - are also illustrated. In this picture space is represented as a two dimensional surface, whereas in full GR each spatial slice is a 3 dimensional volume.}
\label{fig-Foliation}
\end{figure}

\vspace{0.5cm}
\noindent \textbf{Projection operators and spatial slices} \\
Consider the foliation of 4-dimensional spacetime into a 3-dimensional ``spatial'' hyperslice, and a ``timelike'' normal to that slice, as illustrated in figure \ref{fig-Foliation}. It is assumed that the spacetime is \emph{globally hyperbolic}, that is, that it can be foliated into level sets of a universal time function $t$ which are distinct and cover the whole spacetime.

The spatial coordinates $x^i(t)$ label the points on the spatial hypersurface at some coordinate time $t$. Within this slice, the proper distance $dl$ is determined by a 3-dimensional spatial metric $\gamma_{ij}$ according to
\begin{equation}
    dl^2 = \gamma_{ij} dx^i dx^j ~ .
\end{equation}
The normal direction to the hyperslice at each point is given by the unit vector $\vec{n}$, which is the 4-velocity of the normal or Eularian observers. Travelling along this direction, the distance in proper time $\tau$ to the slice at $t+dt$ is given by:
\begin{equation}
    d\tau = \alpha dt \label{eqn:LapseDefinition} ~ .
\end{equation}
Here $\alpha$ is the lapse function, which takes a value at each point on the slice. We can see that the lapse encodes our freedom to foliate the time-like evolution as we choose.

As we move onto the next slice, we may use the equivalent spatial coordinate freedom to relabel the coordinates on our hyperslice. This relabelling is parameterised by the shift vector $\beta^i$ where
\begin{equation}
    x^i(t+dt) = x^i(t) - \beta^i dt \label{eqn:ShiftDefinition} ~ .
\end{equation}
It is probably not immediately be clear why we would want to relabel spatial coordinates - surely it is simpler to leave the points at fixed locations? Unfortunately this is not possible in the general case. Firstly, it turns out that the freedom to move our coordinates dynamically on each slice is important for the stability of our numerical evolution, in particular in black hole spacetimes. Secondly, and more importantly, it is a mistake to think of each coordinate on the spatial slice as a fixed point in space - it is simply a labelling and spacetime can and does stretch between points. Try to define precisely what you mean by ``fixed point'' in a dynamically evolving spacetime, and you'll find that you are in trouble.

Putting these ingredients together, the 4-dimensional spacetime interval $ds$ is given by
\begin{equation}
    ds^2 = (-\alpha^2 + \beta^i \beta_i) dt^2 + 2 \beta_i dx^i dt + \gamma_{ij} dx^i dx^j \label{eqn:ADMMetric} ~ .
\end{equation}
In this adapted basis\footnote{We have, without justifying it, introduced a coordinate system that is adapted to the slicing - the $\vec{e}_0$ basis vector is tangent to the lines of constant $x^i$ (along the $t$ coordinate line), and the $\vec{e}_i$ basis vectors are tangent to the slice. This simplifies things but is not required as we still have 4D diffeomorphism invariance.}, the unit normal vector has the components in raised and lowered forms of
\begin{equation}
    n^\mu = (1/\alpha, - \beta^i/\alpha)  \quad n_\mu = (- \alpha, 0, 0, 0) ~ ,
\end{equation}
from which we can see that it is normalised and timelike such that $n^\mu n_\mu = -1$.

The unit normal vector can be used to define the operator that projects tensors into the spatial hypersurfaces as
\begin{equation}
    P^a_b \equiv \delta^a_b + n^a n_b ~ .
\end{equation}
Applying this to the metric gives the (4-dimensional) metric induced on the spatial slice as
\begin{equation}
    \gamma_{ab} = P^c_a P^d_b g_{cd} = g_{ab} + n_a n_b ~ ,
\end{equation}
from which we can see that the projection operator is in fact the spatial metric, that is $P^b_a = \gamma^b_a$.

\vspace{1cm}
\noindent \textbf{Extrinsic curvature tensor} \\
To fully specify our decomposed spacetime, we must also define an object called the \textit{extrinsic curvature}, denoted by $K_{ab}$. This object describes how the spatial hypersurface is embedded in the 4-dimensional spacetime.

The notion of extrinsic curvature is in some ways more intuitive that the notion of intrinsic curvature. Consider a cylinder in 3-dimensional space - the intrinsic curvature of the 2-dimensional surface is zero - it is flat in the sense that the parallel transport of a vector around a loop on the surface does not lead to a change in its direction. However, our humans brains consider this surface to be curved, which it is \textit{in the 3-dimensions in which it is embedded} - it has an \textit{extrinsic curvature}.

\begin{figure}
\centering
\includegraphics[width=0.5\textwidth]{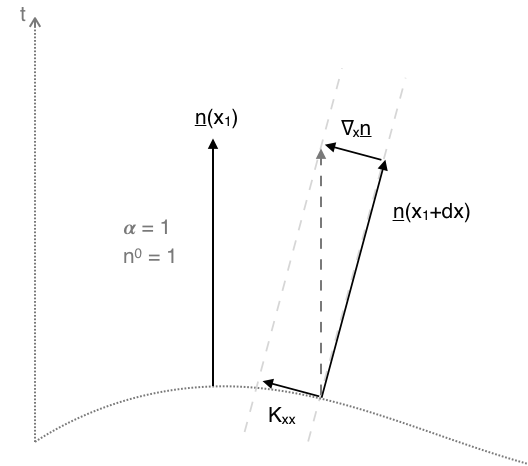}
\caption[Extrinsic Curvature]{In 1+1D the extrinsic curvature in the adapted slicing has only one non trivial component, $K_{xx}$. The figure illustrates how its value relates to the change in the normal vector as it is parallel transported along the slice.}
\label{fig-ExtrinsicCurvature}
\end{figure}

This extrinsic curvature can be defined in two equivalent ways. Firstly, it can be defined as the change in the direction of the normal vector under parallel transport a small distance away along the surface, projected into the surface, as illustrated in Fig. \ref{fig-ExtrinsicCurvature}. That is,
\begin{equation}
    K_{ab} = - P_a^c \nabla_c n_b = - (\nabla_a n_b + n_a n^c \nabla_c n_b) ~ . \label{eqn:Kabdefinition1}
\end{equation}
Equivalently, the extrinsic curvature may be defined as the Lie derivative of the metric along the normal direction, i.e.
\begin{equation}
    K_{ab} = - \frac{1}{2} \pounds_{\vec{n}} \gamma_{ab} ~ .  \label{eqn:Kabdefinition2}
\end{equation}
That this is equivalent to Eq. \ref{eqn:Kabdefinition1} can be shown by expanding out the Lie derivative, which is given as an exercise later on.

Due to the projection, only the spatial components of the extrinsic curvature are non trivial. We can show that
\begin{equation}
    K_{ij} = - \frac{1}{2 \alpha}(\partial_t \gamma_{ij} - D_i \beta_j - D_j \beta_i) ~ . \label{eqn:Kij}
\end{equation}
Where $D_i$ is the covariant derivative defined with respect to the spatial metric $\gamma_{ij}$, or equivalently, the projection of the covariant derivative (and the objects on which it acts) into the spatial slice $D_a \equiv P^b_a \nabla_b$.

Contracting Eq. \eqref{eqn:Kabdefinition1} with the metric it can be seen that its trace is equal to the divergence of the normal lines
\begin{equation}
    K =  g^{ab} K_{ab} = - \nabla_c n^c ~ , \label{eqn:KtraceVolume}
\end{equation}
thus corresponding to the changing volume element of the normal observers. 

\vspace{0.5cm}
\noindent \textbf{Obtaining the equation of motion for the metric using the Einstein equation} \\
In previous sections we have discussed only the kinematics derived from a 3+1 slicing of the metric - everything has been geometric, and there has been no real ``physics'' involved. In this section we derive the dynamics and physical constraints imposed on the metric components by the Einstein equation.

The method to be followed consists in projecting the Einstein equation both onto the spatial surface that we have constructed, and normal to it. In fact there are three options - either both indices can be projected into the spatial hypersurface, both normal to it, or (since the tensors are symmetric), either one of the indices can be projected into the slice and the other one out of it. 

We start from two well-known relations which are simply (but lengthily) derived from the geometric slicing described above (see Gourgoulhon's notes for a full derivation). Firstly, the Gauss-Codazzi equation
\begin{equation}
    P^e_a P^f_b P^g_c P^h_d ~ {}^{(4)}R_{efgh} = {}^{(3)}R_{abcd} + K_{ac} K_{bd} - K_{ad} K_{bc} \, , \label{eqn:GaussCodazzi}
\end{equation}
and secondly the Codazzi-Mainardi equation
\begin{equation}
    P^e_a P^f_b P^g_c n^h ~ {}^{(4)}R_{efgh} = D_b K_{ac} - D_a K_{bc} \, .  \label{eqn:CodazziMainardi}
\end{equation}

Notice how these equations show that the spatial metric and the extrinsic curvature tensor are not entirely independent, for a given 4-metric.

Contracting both sides of equation \eqref{eqn:GaussCodazzi} twice with the metric $g^{ac} g^{bd}$ gives
\begin{equation}
    n^a n^b G_{ab} =  {}^{(3)}R + K^2-K_{ab}K^{ab} \, , \label{eqn:ContractedGC}
\end{equation}
from which, using the Einstein equation to replace $G_{ab}$ with $T_{ab}$, and using the adapted basis, we obtain 
\begin{equation}
    \mathcal{H} = {}^{(3)}R + K^2-K_{ij}K^{ij}-16\pi \rho = 0 \, . \label{eqn:HamiltonianConst}
\end{equation}
where $\rho \equiv n^a n^b T_{ab}$ is the energy density measured by a normal observer. This relation is the \textit{Hamiltonian constraint}. It involves no time derivatives and is independent of the gauge parameters, $\alpha$ and $\beta^i$. It is not, therefore, related to the evolution of the quantities but constrains their relation within a slice. It tells us that we are not free to specify any data we like for the metric and the energy density at some initial (or later) time - the data must satisfy this relation or it will not satisfy the Einstein Equation.
Why does this constraint exist? If I were completely free to choose all my quantities, I could put a very large mass in the centre of my space, and insist that the spacetime around it was completely flat. This is clearly not a valid physical scenario, and we see that it is indeed ruled out by Eq. \eqref{eqn:HamiltonianConst}.

The contraction of Eq. \eqref{eqn:CodazziMainardi} gives the projection of the Einstein equation with one index in and one normal to the slice
\begin{equation}
    P^{ab} n^c G_{bc} = D_b (\gamma^{ab} K - K^{ab}) \, ,
\end{equation}
from which, again using the Einstein equation to eliminate $G_{ab}$, we obtain the \textit{momentum constraints} in the adapted basis
\begin{equation}
    \mathcal{M}^i = D_j (\gamma^{ij} K - K^{ij}) - 8\pi S^i = 0 \, , \label{eqn:MomentumConst}
\end{equation}
where $S^i \equiv - \gamma^{i \mu} n^\nu T_{\mu \nu}$ is the momentum density as measured by normal observers. Again these (three) relations must be satisfied by the data on the each slice if it is to represent a true ``physical'' spacetime. However, as with the Hamiltonian constraint, it gives us no data about how the quantities should evolve in time, save that these relations should continue to be satisfied.

The last option is the projection of both indices of the Einstein tensor into the slice. We will again take a geometric relationship as a starting point, which is the final independent, non trivial projection of the 4-dimensional Riemann tensor,
\begin{equation}
    P^f_b P^h_d \left( n^e n^g ~ {}^{(4)}R_{efgh} \right) =  \pounds_{\vec{n}} K_{bd} - K_{bc} K^c_d + \frac{1}{\alpha} D_b D_d \alpha \, . \label{eqn:3rdproj}
\end{equation}
Starting with Eq. \eqref{eqn:GaussCodazzi} again and contracting with the metric $g^{ac}$ we have also that
\begin{equation}
    P^f_b P^h_d \left( {}^{(4)}R_{fh} + n^e n^g ~ {}^{(4)}R_{efgh} \right) = {}^{(3)}R_{bd} + K K_{bd} - K_{bc} K^c_d \, .
\end{equation}
Before we can equate these two (purely geometric) relations, we want to eliminate the term $P^f_b P^h_d {}^{(4)}R_{fh}$. In deriving the constraints, we have eliminated similar terms by expressing them in terms of $G_{ab}$ and then making a substitution for the EM tensor, thereby introducing the ``physics'' of GR. Whilst we effectively do the same here, a slight difference is that we choose instead to replace $R_{ab}$ directly using the alternative form of the Einstein Equation
\begin{equation}
    R_{ab} = 8 \pi (T_{ab} - \frac{1}{2} g_{ab} T) \, .
\end{equation}
where $T \equiv T^a_a$. Combining these results and expressing them in the adapted basis gives the evolution equation for $K_{ij}$ as  
\begin{multline}
    \partial_t K_{ij} = \beta^k \partial_k K_{ij} + K_{ki} \partial_j \beta^k + K_{kj} \partial_i \beta^k - D_i D_j \alpha \\ + \alpha \left({}^{(3)}R_{ij} + K K_{ij} - 2 K_{ik} K^k_j \right) + 4 \pi \alpha \left( \gamma_{ij}(S - \rho) - 2 S_{ij} \right) \, ,  \label{eqn:dtKij}
\end{multline}
where $S_{ij} \equiv \gamma_{i \mu}\,\gamma_{j \nu} T^{\mu \nu}$ and $S \equiv S^i_i$. Combining this with the definition of $K_{ij}$ in Eq. \eqref{eqn:Kij} above, which can be rearranged to give
\begin{equation}
    \partial_t \gamma_{ij} = - 2\alpha K_{ij} + D_i \beta_j + D_j \beta_i \, , \label{eqn:dtgammaij}
\end{equation}
gives a full set of evolution equations for the spatial metric and the extrinsic curvature.

Note that in the code the variable names and definitions are mostly different to what is shown here. If you look at the file \verb|uservariables.py| you will see some notes to help you interpret the different quantities with respect to the ADM ones.

\subsection{The four ingredients of an NR simulation}

There are roughly speaking four key ingredients one has to consider when running an NR simulation: initial conditions, equations of motion, gauge/stability, and diagnostics. These are briefly summarised below, but you will learn more about each one by completing the exercises with the code in the following section.
\begin{enumerate}
    \item \textbf{Initial conditions} \\
    Specifying the initial data amounts to specifying the 6 components of the spatial metric and the 6 components of the extrinsic curvature at each point on the initial hypersurface, given an initial matter configuration. 
    One problem we face is that these are not independent quantities. The data must satisfy the Hamiltonian and momentum constraints, which are a set of four coupled, elliptic PDEs. These are not (in general) trivial to solve.
    In the exercises, we will use a black hole solution with a simple scalar configuration imposed, for which the constraint satisfying solution is easy to find.

    \item \textbf{Equations of motion} \\
    We derived above the equations of motion for the metric in the ADM formalism. In the \texttt{engrenage} code the matter component is provided by a scalar field $u$ with an equation of motion given by the Klein Gordon equation. It has conjugate momentum $v$ ($v \sim \partial_t u$). In the exercises we will not vary the equations of motion for the metric but simply focus on the scalar field equation. The principle of varying the metric evolution is conceptually the same but simply more complicated in practise.

    \item \textbf{Stability and gauge choice} \\
    As noted above, the ADM equations are not ``well-posed", which in practise means that numerical errors tend to grow in an unbounded way. We therefore need to amend the equations of motion to obtain stable formulations. Well posedness has also been a big issue for modified gravity theories, for which (until recently) well posed formulations were not known. We don't have time to go into this issue in any detail, but I expect you will hear a lot about it in the workshop.
    Another issue is that black holes contain singularities - computers do not like infinity! Singularities of black holes must be managed somehow - in the code we use the so called ``moving-puncture" gauge conditions, Where the lapse and shift variables evolve into a configuration that keeps the coordinate observers at a fixed distance from the singularity\footnote{Remember what I said about fixed distances... this statement is meaningful for stationary spacetimes, and can be considered approximately true for a dynamical one.}. We will see how this works in the exercises.

    \item \textbf{Diagnostics and interpretation of results} \\ 
    One of the hardest aspects of NR is interpreting the results one obtains. It is easy to view a simulation and interpret it in a certain way using one's everyday intuition. Sometimes this works but in general it is a dangerous practice, since what appears to be a physical effect may in fact be due to the dynamical coordinates. Any time you try to think ``globally", you are also likely to run into trouble - your view of the simulation is a kind of ``God view'', where you can see the whole of space at an instant of time. No physical observer can do this. 
    Some standard ways to interpret results are:
    \begin{itemize}
        \item{\textbf{Constraint monitoring}: In general the Hamiltonian and momentum constraints are not enforced on each spatial slice. An essential check of whether the simulation is evolving sensibly is that the constraint violation remains bounded and converges to zero with increasing resolution.}
        \item{\textbf{Finding apparent horizons}: The presence of a black hole event horizon is gauge invariant. However, finding these requires an integration over all time. It is therefore common to instead locate \textit{apparent horizons}, which are marginally trapped surfaces on each spatial slice. Whilst these are local rather than global horizons, if we detect an apparent horizon on a time slice, the singularity theorems tell us that it must lie inside an event horizon. Thus if we detect an apparent horizon we can infer that a black hole has formed, and the area of the apparent horizon provides a lower bound on the black hole mass.}
        \item{\textbf{Gravitational/scalar wave extraction}: One can extract gravitational and scalar waveforms in asympototically flat regions of the spacetime. In \texttt{engrenage} we have no gravitational waves due to the spherical symmetry, but one can extract data about scalar waves as a proxy.}
        \item{\textbf{Mass and momentum fluxes}: One can extract meaningful data from the asymtotically flat regions of the spacetime regarding the mass and momenta of the matter and its fluxes. We will focus on this aspect in the exercises.}
    \end{itemize}
\end{enumerate}

\newpage
\section{Exercises for the course}

\subsection{Some exercises to familiarise yourself with the ADM formalism}

These exercises are optional, but may be useful if you are not familiar with the ADM decomposition. \\

\noindent \textbf{3+1 D split of the metric} \\

The projection operator $P^a_b \equiv \delta^a_b + n^a n_b$ projects tensorial objects into the spatial hypersurface for which $n^a$ is the unit time-like normal vector.

\begin{enumerate}
\item Show that the projection of the normal vector is zero, ie, $P^a_b n^b = 0$
\item Show that if a vector $v^a$ is purely spatial, the projection leaves it unchanged, ie, $P^a_b v^b = v^a$
\end{enumerate}

\noindent The spatial metric $\gamma_{ab}$ can be defined as the projection of the 4D spacetime metric $g_{ab}$ into the slice:
\begin{equation}
\gamma_{ab} \equiv P^c_a P^d_b g_{cd} \,.
\end{equation}

\begin{enumerate}[resume]
\item Show that $P^a_b = \gamma^a_b$, meaning that the spatial metric is itself the projection operator.
\end{enumerate}

\noindent The 3+1D decomposition of the 4D spacetime metric $g_{ab}$ gives the line element
\begin{equation}
ds^2=-\alpha^2\,dt^2+\gamma_{ij}(dx^i + \beta^i\,dt)(dx^j + \beta^j\,dt)\,,
\end{equation}
where $\alpha$ and $\beta^i$ are the lapse and the shift vector respectively (the gauge parameters).

\begin{enumerate}[resume]
\item For the Schwarschild metric in regular Schwarzschild coordinates
\begin{equation}
ds^2= - (1 - 2M/r) dt^2 + (1 - 2M/r)^{-1}dr^2 + r^2 d\Omega^2  \,,
\end{equation}
identify the lapse, shift and spatial metric. What happens to the lapse and the spatial metric at the horizon and the singularity? Why might this cause problems in terms of the 3+1 decomposition and its numerical implementation?
\item Consider the coordinate transformation $r = r' \left( 1 + \frac{M}{2r'} \right)^2$. Show that this gives the Schwarzschild metric in isotropic coordinates, which is often used as the starting point for numerical evolutions of black holes,
\begin{equation}
ds^2= - \left( \frac{1 - M/2r'}{1 + M/2r'} \right)^2 dt^2 + (1 + M/2r')^4 \left(dr'^2 + r'^2 d\Omega^2 \right)  \, .
\end{equation}
Identify the lapse, shift and spatial metric. Now what happens at the event horizon? As $r'$ goes to zero? What does this spacetime look like?
\end{enumerate}

\noindent \textbf{Extrinsic curvature} \\

The extrinsic curvature is defined as
\begin{equation}
K_{ab} = - P^c_a \nabla_c n_b  \,,
\end{equation}

\begin{enumerate} [resume]
\item Show that the above definition of the extrinsic curvature in terms of the changing normal direction is equivalent to a Lie derivative of the spatial metric along the normal direction:
\begin{equation}
K_{ab} = - \frac{1}{2} \pounds_{\vec{n}} \gamma_{ab} \, .
\end{equation}
\end{enumerate}

\begin{enumerate}[resume]
\item What is the extrinsic curvature of the Schwarzschild metric above in normal Schwarszchild and isotropic coordinates?
\item Show that the trace of $K_{ab}$ is equal to the divergence of the normal lines
\begin{equation}
K =  g^{ab} K_{ab} = - \nabla_c n^c \, .
\end{equation}
\end{enumerate}

\noindent \textbf{ADM equations} \\

We derived the Hamiltonian constraint above. The (three) momentum constraint(s) will be done here and you are encouraged to try the evolution equation for the extrinsic curvature $K_{ij}$ if you would like more practise.

\begin{enumerate}[resume]
\item Starting from the Codazzi-Mainardi relation
\begin{equation}
P^e_a P^f_b P^g_c n^h ~ {}^{(4)}R_{efgh} = D_b K_{ac} - D_a K_{bc} \, ,
\end{equation}
where $D_a \equiv P^b_a \nabla_b$, derive the momentum constraint
\begin{equation}
\mathcal{M}^a \equiv D_b (\gamma^{ab} K - K^{ab}) - 8\pi S^a = 0 \,
\end{equation}
where $S^a \equiv - \gamma^{ab} n^c T_{bc}$
\end{enumerate}

\subsection{Exercises using the code \texttt{engrenage}}

There are more detailed set up instructions in the wiki at: \\
\url{https://github.com/GRTLCollaboration/engrenage/wiki}\\
but the description of the exercises below is designed to be mostly self contained. \\

\noindent You may want to look at the pdf lecture notes in the wiki at:\\
\url{https://github.com/user-attachments/files/16046298/LisbonJuly2024.pdf}\\
for more details of how the engrenage variables relate to the ADM decomposition described above. The formalism is that of \cite{Ruchlin:2017com,Baumgarte:2012xy,Brown:2009dd}, which differs somewhat from the standard BSSN treatment. However, most of the variables you will look at in these exercises -- the lapse $\alpha$, shift $\beta^r$, mean curvature $K$, scalar field $u$ and its conjugate momentum $v$, are the same in most formulations. \\

\noindent \textbf {Set up}

\noindent You should use the following link to access the code in the NewHorizonsForPsi branch: \\
\url{https://github.com/GRTLCollaboration/engrenage/tree/NewHorizonsForPsi}. 

Once there you can either fork and clone using git if you know how or (which I recommend) click on the big green ``Code'' button and click \verb|Download Zip| to get the whole code in a zipped folder. Unzip it.

I assume you already have a python3 installation such as anaconda, and can open a terminal and navigate to the folder (if not ask me!). If using \verb|conda| you should install the packages in a virtual environment by running 
\begin{verbatim}
conda create -n myenv jupyterlab notebook matplotlib numpy scipy tqdm sympy
\end{verbatim}
and activate it using 
\begin{verbatim}
conda activate myenv
\end{verbatim}
(or use \verb|pip| if you prefer as in the \verb|README.md| file).
Then run
\begin{verbatim}
jupyter notebook
\end{verbatim}
to open an interface in the web browser where you can view and run the files. At the end of the session run
\begin{verbatim}
conda deactivate
\end{verbatim}
to deactivate the environment. 

\vspace{0.5cm}
\noindent \textbf {Exercise 1: Initial conditions}

In this exercise you will add a scalar field to the black hole, and observe its evolution from a homogeneous initial state to a (quasi) stationary configuration that is related to the confluent Heun functions that appear in many aspects of black hole perturbation theory.

Steps:
\begin{itemize}
    \item First make a copy of the black hole example notebook \verb|BHEvolution.ipynb| and the initial conditions for the black hole in \verb|bhinitialconditions.py|. (It is bad practise to amend existing examples, even if you are using git to track your changes, and much better to keep the existing example for reference.)
    \item Now add a small spatially constant scalar field $u_0 = 10^{-6}$ to the black hole initial conditions. What density does this create and how does it compare to the density (as characterised by the curvature radius) of the black hole?
    \item We need to make sure the Hamiltonian constraint is still solved, so also set the extrinsic curvature $K$ to achieve this.
    \item You should see the field evolve into an oscillating configuration, similar to the plots in \url{https://arxiv.org/pdf/1904.12783}. What limits how long you can meaningfully run this simulation?
\end{itemize}

\vspace{0.5cm}
\noindent \textbf {Exercise 2: Change the scalar equation of motion} 

In this exercise you will change the scalar field mass and add self interactions in the field, and observe how it changes the scalar profiles.

Steps:
\begin{itemize}
    \item Find where the potential of the scalar field is set. 
    \item Add a self interaction term with the form 
    \begin{equation}
        V(u) = \frac{1}{2}\mu^2 u^2 + \frac{1}{4} \mu^2 \lambda u^4
    \end{equation}
    \item Investigate the effect of changing the scalar mass and the self interaction.
    \item For the mass dependence, you should see the profile change as illustrated in Fig 1. of \url{https://arxiv.org/pdf/1904.12783}, with a larger mass giving shorter wavelength oscillations. What limits the mass ranges you can study?
    \item For the dependence on $\lambda$, a repulsive coupling (positive lambda) should result in the field having a smaller amplitude than the massive case and saturating at some value. The attractive (negative) coupling should result in a larger amplitude.
    \end{itemize}

\vspace{0.5cm}
\noindent \textbf {Exercise 3: Implement a shock avoiding gauge} 

In this exercise you will amend the evolution of the lapse to implement the shock avoiding gauge recently used in \cite{Baumgarte:2022ecu}.

Steps:
\begin{itemize}
    \item Set the mass to 1.0 and self interaction to zero. Rerun the example and plot the evolution of the lapse and shift over time. You should see the ``collapse of the lapse'' where the proper time of the normal observers slows near to the singularity. The shift is positive near the black hole, which moves the coordinate observers away from the singularity, relative to the normal observers who are infalling.
    \item Find and amend the evolution equation for the lapse to implement the new condition where:
    \begin{equation}
        \partial_\tau \alpha = -(\alpha^2 + \kappa) ~K ~.
    \end{equation}
    Here $\tau$ is the proper time of the normal observers. It may help to note that currently the code uses the standard 1 + log slicing condition
    \begin{equation}
        \partial_\tau \alpha = -2 \alpha K ~.
    \end{equation}
    \item Initially I recommend you set $\kappa = 0.05$ and investigate the effect of this change on the evolution of the lapse. What do you observe? What does this mean? Is the stability sensitive to the coefficient chosen?
    \item If you like, try turning off the evolution of the lapse (i.e. just set the time derivative to zero) and verify that it doesn't work to have a non dynamical lapse!
\end{itemize}

\vspace{0.5cm}
\noindent \textbf {Exercise 4: Diagnostics - matter flux} \\

In this exercise you will implement a function to evaluate the matter fluxes at each radius. 

Steps:
\begin{itemize}
    \item Create a new diagnostic file - I suggest basing it on the Hamiltonian constraint diagnostic at \verb|hamdiagnostic.py|.
    \item Implement the equation for the flux measured by a normal observer for the scalar field at each radius, which you can show is
    \begin{equation}
        F(r) = 4\pi r^2 \sqrt{\gamma} S^r 
    \end{equation}
    with $S_i = v ~\partial_j u$. In the form of the variables used, $\gamma = r^2 e^{12\phi}$ and $\gamma^{ij} = e^{-4\phi}\bar{\gamma}^{ij}$.
    \item What happens to the flux at small radii over time? Making the mass larger may help to see this easier.
    \item You can also implement a diagnostic for the total mass of the matter within the surface. Is the change in the total mass equal to the flux? If not why not and what is the missing term to reconcile the two? 
\end{itemize}

\vspace{0.5cm}
\noindent \textbf {Extensions} \\
In all these exercises the backreaction on the metric of the scalar was small. Try turning up the amplitude of the scalar and looking at the impact on the conformal factor $\phi$.

You may also want to repeat the exercises, in particular adding a self interaction to the oscillaton example, where the balance between the metric and scalar is stronger and an attractive interaction can destabilise it and cause it to form a black hole.

\newpage
\section{Hints for exercises}

\subsection{Some exercises to familiarise yourself with the ADM formalism}

Here are some hints for the problems:

\begin{itemize}
\item The normal is a unit timelike vector, so $n^a n_a = -1$.
\item If $v^a$ lies in the slice its dot product with the normal vector $n_a v^a$ will be zero.
\item You can expand out the spatial metric as $\gamma_{ab} = g_{ab} + n_a n_b$
\item You can expand out the 4-metric as $g_{ab} = \gamma_{ab} - n_a n_b$
\item The unit normal vector is orthogonal to its gradient, so $n_a \nabla_b n^a = 0$
\item The metric is compatible with the (metric compatible) covariant derivative $\nabla_a g_{bc} = 0$
\item One can show that the partial derivatives in the Lie derivative
\begin{equation}
\pounds_{\vec{V}} T^a_b = V^c \partial_c T^a_b - T^c_b \partial_c V^a + T^a_c \partial_b V^c \ ,
\end{equation}
can be replaced by the metric compatible covariant derivative with the same result, that is
\begin{equation}
\pounds_{\vec{V}} T^a_b = V^c \nabla_c T^a_b - T^c_b \nabla_c V^a + T^a_c \nabla_b V^c \, .
\end{equation}
\item For (9) start by contracting both sides twice with the metric $g^{ac}g^{bd}$, and use the Einstein equation to replace the Einstein tensor $G_{ab}$. It may also help to show that $P^{ab} n^c G_{bc} = P^{ab} n^c R_{bc}$
\end{itemize}

\subsection{Exercises using the code \texttt{engrenage}}

\noindent \textbf {Exercise 1: Initial conditions}

Look at how the lapse is being set. Can you do something similar for the field $u$ and $K$?

Don't worry about constraint violation around $r~0$ as this relates to the puncture gauge. Assume that the BH data already satisfies the Hamiltonian constraint. Then the addition of the scalar field only adds a spatially constant term $\rho = V(u_0)$. 

You can use Eq. (13) from the Baumgarte paper in the \verb|papers| folder. Why is Eq. (14) not needed?

In practise the addition of the field and $K$ won't make a huge difference as long as the field value is small, as the error is dominated by the spatial gradients in the metric.

You may need to adjust the limits on the plots to be able to see the oscillations.

\vspace{0.5cm}
\noindent \textbf {Exercise 2: Change the scalar equation of motion}

The file you need to edit is \verb|mymatter.py|.

After changing the mass and interaction values you may need to manually restart the kernel before rerunning for the changes to take effect.

It may be easier to see the difference if you plot the evolution of specific radial values rather than the profiles. You may also have to make the values bigger, or the timescales longer. 

It is best to stick to roughly order 1 numbers for the mass, as otherwise the timescales of oscillation differs a lot from the BH crossing time.

Beware that attractive interactions may become unstable if the field value gets large enough to reach the part where the potential is unbounded!

\vspace{0.5cm}
\noindent \textbf {Exercise 3: Implement a shock avoiding gauge} 

The file you need to edit is \verb|rhsevolution.py|.

More details on the gauge and choice of kappa are given in \url{https://inspirehep.net/literature/2111279}.

Instability usually means that the solver gets stuck, it doesn't always quit by itself. If each timestep is taking ages, probably something is wrong and you should interrupt the kernel. 

If you find instability, try reducing the maximum time so that you can see what is going wrong in the variables before it gets stuck.

\vspace{0.5cm}
\noindent \textbf {Exercise 4: Diagnostics - matter flux}

Remember to raise the index on $S_r$ using the metric.

If you get stuck I suggest you take a quick look at the solution for inspiration and then try to recreate it without looking at it again!

This paper explains why the fluxes don't balance the mass and gives a more detailed discussion of how fluxes are defined and their slicing dependence \cite{Clough:2021qlv}.

\section{Solutions to exercises}

Solutions to the exercises are provided on the workshop wiki page here: \\
\url{https://github.com/GRTLCollaboration/engrenage/wiki/New-Horizons-for-Psi-workshop}.

\newpage
\section{Acknowledgements}

KC thanks the organisers of the New Horizons workshop for their invitation to teach this course, for acting as welcoming hosts in the beautiful city of Lisbon, and for the interesting discussions that were generated at the workshop. She acknowledges funding from the UKRI Ernest Rutherford Fellowship (grant number ST/V003240/1) and STFC Research Grant ST/X000931/1 (Astronomy at Queen Mary 2023-2026).
The \texttt{engrenage} code is based on a spherically adapted code that was generously shared by Thomas Baumgarte, and the NRpy code of Zac Etienne, in particular the formalism used closely follows that described in the papers \cite{Ruchlin:2017com,Baumgarte:2012xy,Brown:2009dd}.
The \texttt{engrenage} code has also benefitted from input from Nils Vu, Leo Stein, Cristian Joana, Cheng-Hsin Cheng, David Sola Gil and Marcus Hatton. Since the workshop, the code has been further improved by Maxence G\'erard (\'Ecole Polytechnique), and Josu Aurrekoetxea.

\bibliographystyle{unsrt}
\bibliography{mybib}

\end{document}